\documentstyle[12pt]{article}
\def\Journal#1#2#3#4{{#1} {\bf #2},#3, (#4)}
\def\NPA{{\em Nucl.Phys.} A}

\def\PRL{\em Phys.Rev.Lett.}
\def\PRC{{\em Phys.Rev.} C}

\def\PLB{{\em Phys.Lett.} B}

\def\PTP{\em Prog.Theor.Phys.}
\begin{document}
\begin{titlepage}


\vspace{1cm}

\centerline{\large \bf $N\Omega$ AND $\Delta\Omega$ DIBARYONS IN $SU(3)$ CHIRAL QUARK MODEL
\footnote{This work is supported in part by the National Natural Science
Foundation of China under the grant number 1977551B and the Chinese
Academy of Sciences under the grant number B78.}}

\vspace{1cm}

\centerline{Q.B.Li${^a}$, P.N.Shen$^{a,b,c,d}$}

\vspace{1cm}

{\small
{
\flushleft{\bf  $~~~$a.Institute of High Energy Physics,Chinese Academy
of Sciences,}
\vspace{-0.2cm}
\flushleft{\bf  $~~~~~$P.O.Box 918(4), Beijing  100039, China}
\flushleft{\bf  $~~~$b.Chinese Center of Advanced Science and Technology,
(World }
\vspace{-0.2cm}
\flushleft{\bf  $~~~~~$Laboratory), P.O.Box 8730, Beijing 100080, China}
\flushleft{\bf  $~~~$c.Institute of Theoretical Physics,Chinese Acedemy
of Sciences,}
\vspace{-0.2cm}
\flushleft{\bf  $~~~~~$P.O.Box 2735, Beijing 100080, China}
\flushleft{\bf  $~~~d.$ Center of Theoretical Nuclear Physic, National Lab
of Heavy Ion}
\vspace{-0.2cm}
\flushleft{\bf  $~~~~~$Accelerator, Lanzhou 730000, China}
}}

\vspace{1cm}

\centerline{\bf Abstract}

\baselineskip 30pt
\noindent
      The binding energy of the six quark system with strangeness
 s=-3 is investigated under the chiral $SU(3)$ constituent quark
 model in the framework of $RGM$. The calculations of the single
 $N\Omega$ channel with spin S=2 and the single $\Delta\Omega$ channel
 with spin S=3 are performed. The results show that both
 systems could be dibaryons and the interaction induced by the
 chiral field  plays a very important role on forming
 bound states in the systems considered.
 The phase shifts and scattering lengths in corresponding channels are
 also given.
\end{titlepage}

\baselineskip 30pt

\noindent
{\bf {\large 1. Introduction}}

\vspace{0.3cm}

Dibaryon, which contains six quarks, has been an important object
to be investigated both theoretically and experimentally since Jaffe
predicted the H particle in 1977 [1]. Different from the deuteron,
whose property can be well explained by the meson exchange mechanism
on the baryon level, this object is supposed to be a color-singlet
multi-quark system with a sufficiently smaller size, and the quark-gluon
degrees of freedom could be dominant degrees. 
No doubt, the dibaryon should be
a perspective field to study the phenomenology of the strong interaction.
Through study, we can enrich our knowledge of the strong interaction
between quarks in the short-range region and further understand
the basic theory of the strong interaction, Quantum Chromodynamics
$(QCD)$, especially its nonperturbative effect.

\vspace{0.3cm}

In past twenty years, various kinds of dibaryons were proposed.
Among these particles, some of them are strangenessless such as $d^*$
[2,3], $d'$ [4] and etc., and the others carry strangeness[2,5].
The H particle is just the one that has most intensively been investigated
in both theoretical and experimental studies. The advantage of investigating
dibaryon with strangeness is that one can study the strong interaction
from a variety of sides by using the new data involving strangeness
and proposing new model theories and learn how to properly
deal with the nonperturbative $QCD$ (NPQCD) effect and further the
hadronic structure. Recently, it was reported that in the system with
multi-strangeness, the dibaryon state is highly possible to show up [5].
Yu et al. predicted the possible existences of $\Omega\Omega(S=0,L=0)$
and $\Xi\Omega(S=1,L=0)$ (single channel calculation) dibaryons [5],
where S and L denote the spin and angular momentum, respectively.
Li and shen announced the possible existences of $\Xi^*\Omega (S=0)$
and $\Xi\Omega$-$\Xi^*\Omega (S=1,L=0)$ (coupled channel calculation)[6,7].
In this paper, we would study the possible S-wave dibaryons in the
system with s=-3, namely, the $N\Omega$ (S=2,T=1/2,L=0) and
$\Delta\Omega (S=3,T=3/2,L=0)$ dibaryons.
Because of the available nucleon beam and the only weak decay modes of
$\Omega$, it would be easier to find the $N\Omega$ dibaryon than the
others such as H and $d^*$. Experimentally searching $N\Omega$ would be
an especially interesting investigation.

\vspace{0.3cm}

In the next section, the model employed for this study is briefly given.
In Section 3, the results are presented and discussed. And the conclusion
is drawn in Section 4.

\vspace{0.5cm}

\noindent
{\bf {\large 2. Model}}

\vspace{0.3cm}

Nowadays, more  evidences show that $QCD$
is the underlying theory of the strong interaction. However, because of the
complexity of the nonperturbative effect of $QCD$ at the lower energy region,
one has to develop $QCD$-inspired models. In early 1980's, the
constituent quark model showed it's power in studying nucleon-nucleon ($N-N$)
interactions[8]. By considering the coupling between the constituent
quark and chiral fields, that model was modified to the chiral quark model
[9], and later, was further extended to the SU(3) chiral-quark model
[10]. In the SU(3) chiral quark-model, both pseudoscalar nonet mesons
$\pi~,~K~,~\eta$ and $\eta'$ and the scalar nonet mesons $\sigma~,~\sigma'~,
~\kappa$ and $\epsilon$ are involved in the interaction between quarks.
With such model, not only the single baryon properties can be basically
explained [11], but also the scattering data of the $N-N$ and the
nucleon-hyperon ($Y-N$) processes can be well-reproduced [10].
Applying this model to the system with two strangeness without further
introducing additional parameters, the resultant
binding energy of the H particle is consistent with the experimental
data available [12,13]. These are just the basic requirements for a model
which would be employed to study the nuclear systems where no experimental
data are available and to predict their properties. In this paper, we would
use the chiral $SU(3)$ quark model to study the bound state
problem in the systems with s=-3 and believe that the results would be
reasonable and reliable.

\vspace{0.3cm}

In the chiral quark model, the dynamics of the dibaryon, the
six-quark system, is governed by the Hamiltonian
\begin {eqnarray}
          H~=~\sum_{i}T_i~-~T_G~+~\sum_{i<j}V_{ij} ,
\end{eqnarray}
where the first and second terms are the total kinetic energy operator of
the system and the kinetic energy operator of the center of mass motion
(CM) of the system, respectively. $V_{ij}$ represents the interaction
between the $i$-th and $j$-th constituent quarks and can be written as
\begin{eqnarray}
       V_{ij}~=~\sum_{i<j}(V^{conf}_{ij}~+~V^{ch}_{ij}
            ~+~V^{OGE}_{ij})  .
\end{eqnarray}
In this equation, the confinement potential $V^{conf}_{ij}$ describes
the long range effect of $NPQCD$, the one-gluon exchange $(OGE)$
potential $V^{OGE}_{ij}$ mainly depicts the short-range
perturbation $QCD$ $(PQCD)$ effect. The forms of these potentials can be
found in Ref.[10]. $V^{ch}_{ij}$ denotes the chiral-quark field coupling
induced potential
\begin{eqnarray}
       V^{ch}_{ij}~=~\sum_{a}V_{\pi_a}({\bf r}_{ij})
                ~+~\sum_{a}V_{\sigma_a}({\bf r}_{ij}) ,
\end{eqnarray}
where subscripts $\pi_a$ and $\sigma_a$ can be
pseudoscalar mesons
$\pi~,~K~,~\eta$ and $\eta'$ and scalar mesons $\sigma~,~\sigma'~,~\kappa$
and $\epsilon$, respectively.
The explicit forms of these potentials can also be found in Ref.[10].
The wave function of the single baryon can be expressed as:

\begin{eqnarray}
\setlength{\unitlength}{0.33cm}
\begin{picture}(40,11)(4,0)
\multiput(6,9)(0,1){3}{\framebox(1,1)}
\put(11,10){$=$}
\multiput(12.5,10)(1,0){3}{\framebox(1,1)}
\put(18.5,10){$\times$}
\multiput(20.5,9)(0,1){3}{\framebox(1,1)}
\put(7,8){$OFSC$}
\put(15.5,9){$OFS$}
\put(22,8){$c$}
\put(23,10){$For~deculplet~baryons~(S=3/2)$}
\put(11,5){$=$}
\put(12,5){$\frac{1}{\sqrt{2}}$}
\put(14,5){${\bf(}$}
\multiput(15,5)(1,0){2}{\framebox(1,1)}
\put(15,4){\framebox(1,1)}
\put(17,6){$MS$}
\put(17,3){$OF$}
\put(19,5){$\times$}
\multiput(20,5)(1,0){2}{\framebox(1,1)}
\put(20,4){\framebox(1,1)}
\put(22,6){$MS$}
\put(22,3){$S$}
\put(24,5){$+$}
\multiput(25,5)(1,0){2}{\framebox(1,1)}
\put(25,4){\framebox(1,1)}
\put(27,6){$MA$}
\put(27,3){$OF$}
\put(29,5){$\times$}
\multiput(30,5)(1,0){2}{\framebox(1,1)}
\put(30,4){\framebox(1,1)}
\put(32,6){$MA$}
\put(32,3){$S$}
\put(34,5){${\bf)}$}
\put(35,5){$\times$}
\multiput(36,3)(0,1){3}{\framebox(1,1)}
\put(37,2){$c$}
\put(23,0){$For~octet~baryons~(S=1/2).$}
\end{picture}
\end{eqnarray}
It should be mentioned that due to the flavor symmetry breaking,
the wavefunctions in the orbit and flavor spaces are always associated.
The quark wave function of specific type $i$ ($i$=up, down or strange)
in a baryon, can be written as
\begin{eqnarray}
    \Phi_i({\bf r}_i)~=~(1/{\pi}b^{2}_{i})^{3/4}exp[-({\bf r}_i-{\bf R})^2/2b^{2}_{i}] ,
\end{eqnarray}
where $\bf R$ is the coordinate vector of the center of mass motion of the
baryon. The width parameter $b_i$ is associated with the oscillator frequency
$\omega$ by the constituent quark mass $m_i$
\begin{eqnarray}
   \frac{1}{b^{2}_{i}}~=~m_i\omega.
\end{eqnarray}
Then the wave function of the dibaryon with quantum numbers $S$ and $T$,
which should be totally antisymmetric, can be written in the framework
of the Resonating Group Method $(RGM)$ as
\begin{eqnarray}
  {\Psi}_{6q}~=~{\cal A}[{\Phi}_A{\Phi}_B\chi({\bf R}_{AB})Z({\bf R}_{CM})]
\end{eqnarray}
where ${\chi}({\bf R})$ is the trial relative wave function between
the clusters A and B, respectively, $Z({\bf R}_{CM})$ represents the
CM  wave function of the six quark system and ${\cal A}$ denotes the
antisymmetrizer. Expanding unknown ${\chi}({\bf R})$ by well-defined
basis wavefunctions, such as Gaussian functions, one can solve RGM bound
state equation to obtain eigenvalues and corresponding wave
functions, simultaneously. The details  of solving RGM bound state problem
can be found in Refs.[14,15].

\vspace{0.3cm}

All the model parameters employed in the calculation are fixed
by the mass splittings among $N$, $\Delta$, $\Lambda$, $\Sigma$ and $\Xi$,
respectively, and the stability conditions of the octet $(S=1/2)$ and
decuplet $(S=3/2)$ baryons, respectively.

\vspace{0.3cm}

We used two sets of parameters to calculate the binding energy of the
$N\Omega (S=2,T=1/2)$ and $\Delta\Omega (S=3,T=3/2)$ systems
in the $SU(3)$ chiral quark model. The first set of parameters has widely
been used in our previous investigations[10,11,5]. Because the strange
clouds around baryons may be influential for a system with s=-3, in Set 2,
we further choose the masses of $\kappa$ and $\epsilon$ mesons to be those
which have same quantum numbers in the Particle Data Table (PDT)[16].
The values of these parameters are tabulated in Table 1.

\centerline{\bf {Table 1~~~~Model parameters under $SU(3)$ chiral quark model}}
{\small
\begin{center}
\begin{tabular}{ccc|ccc}\hline
                              & $~~~~~$    Set1  & $~~~~~$  Set2 &                                   & $~~~~~$  Set1  & $~~~~~$  Set2    \\\hline
  $m_u~(MeV)$                 & $~~~~~$    313   & $~~~~~$  313  &                                   &                &                  \\
  $m_s~(MeV)$                 & $~~~~~$    470   & $~~~~~$  470  &                                   &                &                  \\
  $b_u~(fm)$                  & $~~~~~$    0.505 & $~~~~~$  0.505 &                                   &                &                  \\
  $m_\pi~(fm^{-1})$           & $~~~~~$    0.7   & $~~~~~$  0.7  & $\Lambda_\pi~(fm^{-1})$           & $~~~~~$  4.2   & $~~~~~$  4.2     \\
  $m_K~(fm^{-1})$             & $~~~~~$    2.51  & $~~~~~$  2.51 & $\Lambda_k~(fm^{-1})$             & $~~~~~$  4.2   & $~~~~~$  4.2     \\
  $m_\eta~(fm^{-1})$          & $~~~~~$    2.78  & $~~~~~$  2.78 & $\Lambda_\eta~(fm^{-1})$          & $~~~~~$  5.0   & $~~~~~$  5.0     \\
  $m_{\eta'}~(fm^{-1})$       & $~~~~~$    4.85  & $~~~~~$  4.85 & $\Lambda_{\eta'}~(fm^{-1})$       & $~~~~~$  5.0   & $~~~~~$  5.0     \\
  $m_\sigma~(fm^{-1})$        & $~~~~~$    3.17  & $~~~~~$  3.17 & $\Lambda_\sigma~(fm^{-1})$        & $~~~~~$  4.2   & $~~~~~$  7.0     \\
  $m_{\sigma'}~(fm^{-1})$     & $~~~~~$    4.85  & $~~~~~$  4.85 & $\Lambda_{\sigma'}~(fm^{-1})$     & $~~~~~$  5.0   & $~~~~~$  5.0     \\
  $m_\kappa~(fm^{-1})$        & $~~~~~$    4.85  & $~~~~~$  7.09 & $\Lambda_\kappa~(fm^{-1})$         & $~~~~~$  5.0   & $~~~~~$  7.61     \\
  $m_\epsilon~(fm^{-1})$      & $~~~~~$    4.85  & $~~~~~$  7.09 & $\Lambda_\epsilon~(fm^{-1})$      & $~~~~~$  5.0   & $~~~~~$  7.61     \\
  $g_u$                       & $~~~~~$  0.936   & $~~~~~$ 0.936 &                                   &                &                  \\
  $g_s$                       & $~~~~~$  0.924   & $~~~~~$ 0.781 &                                   &                &                  \\
  $a_{uu}~(MeV/fm^2)$         & $~~~~~$  54.34   & $~~~~~$ 57.71 & $a^0_{uu}~(MeV)$                  & $~~~~~$ -47.69 & $~~~~~$ -48.89   \\
  $a_{us}~(MeV/fm^2)$         & $~~~~~$  65.75   & $~~~~~$ 66.51 & $a^0_{us}~(MeV)$                  & $~~~~~$ -41.73 & $~~~~~$ -50.57   \\
  $a_{ss}~(MeV/fm^2)$         & $~~~~~$  102.97  & $~~~~~$ 115.39 & $a^0_{ss}~(MeV)$                  & $~~~~~$ -45.04 & $~~~~~$ -68.11   \\\hline               &

\end{tabular}
\end{center}
}
In this table, $m_A$ denotes the mass of particle $A$ which can be
either the valence quark or the meson involved, $\Lambda_A$ represents
the corresponding cut-off mass of the particle, $g_q$ is the
coupling constant between gluon and valence quark q,
$a_{q_{1}q_{2}}$ depicts the confinement strength between valence
quarks $q_1$ and $q_2$ and
$a^0_{q_{1}q_{2}}$ is the corresponding zero-point energy in the confinement
potential, respectively. It should be mentioned that by using
these two sets of parameters, we can meet the above mentioned
requirement, namely, the mentioned empirical data can be well
reproduced. A typical example can be seen in Fig.1, where the
solid and dashed curves represent the results by using Sets 1
and 2, respectively.

\vspace{0.5cm}

\noindent
{\bf {\large 3. Results and discussions}}

The resultant binding energies and corresponding root-mean-square
radii (RMS) of the systems concerned are tabulated in Table 2.
It is shown that with Set 1 the binding energies of the
$N\Omega (S=2,T=1/2)$ and $\Delta\Omega (S=3,T=3/2)$ systems are
3.5 MeV and 4.4 MeV, respectively, and the corresponding RMS' are
1.18 fm and 1.15fm, respectively. With Set 2, the corresponding
binding energies are 12.7 MeV and 14.2 MeV for $N\Omega$ and
$\Delta\Omega$, respectively, and the corresponding RMS' are
0.98 fm and 0.96 fm, respectively. Deep binding behavior in Set 2
indicates that the strange cloud has the positive effect in forming
dibaryons in the $N\Omega (S=2,T=1/2)$ and $\Delta\Omega (s=3,T=3/2)$
systems in the chiral $SU(3)$ quark model.

\centerline{\bf {Table 2~~~~Binding energy $B_{AB}$ and RMS}}
\centerline{\bf {{of $N\Omega (S=2,T=1/2)$ and
$\Delta\Omega (S=3,T=3/2)$}$^{\dag}$}}

\begin{center}
\begin{tabular}{|c|c|c|c|c|}
\hline
$~$          & \multicolumn{2}{|c|}{one channel}
& \multicolumn{2}{|c|}{one channel} \\
Channel  & \multicolumn{2}{|c|}{$N\Omega (S=2,T=1/2)$}
 & \multicolumn{2}{|c|}{$\Delta\Omega (S=3,T=3/2)$} \\
\cline{2-5}
  & $B_{N\Omega}$ & $RMS$ & $B_{\Delta\Omega}$ & $RMS$ \\
&$(MeV)$ & $(fm)$ & $(MeV)$ & $(fm)$   \\  \hline
$SU(3)$ Set1 &  3.5 & 1.18  &  4.4 & 1.15  \\  \hline
$SU(3)$ Set2 &  12.7 & 0.98  &  14.2 & 0.96  \\  \hline
ext. $SU(2)$ &  31.8 & 0.81   &  34.3 & 0.80  \\ \hline
  $SU(2)$    &  49.5 & 0.74   &  49.5 & 0.74  \\ \hline
\end{tabular}
\end{center}
$\dag$ {\footnotesize {$B_{AB}$ denotes the binding energy between
A and B baryons and RMS represents the corresponding root-mean-square
radius.}}

\vspace{0.3cm}

It should be emphasized that in these two systems, the expectation values
of the antisymmetrizer in the spin-flavor-color space is equal to 1.
It means that there is no quark exchange effect, and whether the system
is bound merely depends on the characteristics of the interaction induced
by the chiral-quark field coupling. To further demonstrate the effects of
chiral clouds, we study the $N\Omega (S=2,T=1/2)$ and
$\Delta\Omega(S=3,T=3/2)$ systems by employing other
two chiral quark models, the extended $SU(2)$ model and the
$SU(2)$ model. Comparing with the $SU(3)$ chiral quark model,
in the so-called extended $SU(2)$ chiral quark model,
the scalar mesons $\sigma'$, $\kappa$, and $\epsilon$ are eliminated,
while in the $SU(2)$ chiral quark model, only pseudoscalar meson
$\pi$ and scalar meson $\sigma$ are remained.
The results are also tabulated in Table 2. It shows that
the binding energies of these two systems in the extended
$SU(2)$ chiral quark model and $SU(2)$ chiral quark model are larger
than those in the $SU(3)$ chiral quark model and these systems are
even bounder in the $SU(2)$ chiral quark model. The reason is clear.
Because ignoring the repulsive-natured $\epsilon$, the relative
wavefunction would distribute more inward. As a consequence, the
expectation value of the $\sigma$ meson induced interaction would pick
up more strength from the $\sigma$ induced potential so that more
attractive character presents. For further reference, we give the
energy level sketches of these systems in Figs. 2 and 3. The shaded
areas are our predictions.

\vspace{0.3cm}

To cross-check our predictions,
we demonstrate the S-wave phase shifts of the $N\Omega(S=2,T=1/2)$ and the
$\Delta\Omega(S=3,T=3/2)$ systems in the $SU(3)$ chiral quark model with
parameter Sets 1 and 2 in Fig.4 and 5, respectively.
In these figures, solid curves are the results with Set 1 and the dashed
curves represent the results with Set 2. According to the phase shifts,
one can estimate the scattering length $a$. The results are
presented in Table 3:

\centerline{\bf {Table 3~~~~The Scattering length $a$ of the dibaryons}}
\begin{center}
\begin{tabular}{|c|c|c|}
\hline
        &   one channel          &     one channel\\
        &   $N\Omega~(S=2)$  & $\Delta\Omega~(S=3)$\\ \hline
  Set 1 &   $-3.68~(fm)$         &     $-3.40~(fm)$\\ \hline
   Set 2 &   $-2.32~(fm)$         &     $-2.22~(fm)$\\ \hline
\end{tabular}
\end{center}

Both phase shifts and scattering lengths are consistent with our
findings.

\vspace{0.5cm}

\noindent
{\bf {\large 4. Conclusion}}

In the framework of the $SU(3)$ chiral quark model, we investigated  the
possible S-wave baryon-baryon bound states in the s=-3 sector of the six-quark
system. By employing either the $SU(3)$ or extended $SU(2)$ or $SU(2)$
chiral quark models, we would report that there may exist two bound states,
$N\Omega$ (S=2,T=1/2,L=0) and $\Delta\Omega (S=3,T=3/2,L=0)$. The results
show that these systems are bounder in the $SU(2)$ chiral quark model and less
bound in the $SU(3)$ chiral quark model. If we believe that the $SU(3)$ chiral
quark model is more correct, the resultant binding energies of
$N\Omega$ (S=2,T=1/2,L=0) and $\Delta\Omega (S=3,T=3/2,L=0)$ should be ranged
from 3.5 to 12.7 MeV and from 4.4 to 14.2 MeV, respectively, and their RMS' are
$1.18\sim 0.98$ fm and $1.15\sim 0.96$ fm, accordingly.
The predictions in the extended $SU(2)$ and $SU(2)$ chiral quark model
can serve as alternative references for searching $N\Omega$ and
$\Delta\Omega$. Because the binding energy of the $\Delta\Omega$ system in
any model are above the threshold of the $N\Omega\pi$ channel which is due to
the only strong decay modes of $\Delta$, $\Delta\rightarrow$ $N\pi$,
$\Delta\Omega$ would have a broad width and might not be easy to detect
in the experiment. However, $\Omega$ baryon can only decay through the weak
mode, $N\Omega$ would have a narrow width. Considering the possible 
nucleon beam, it would especially be interesting to search
$N\Omega (S=2,T=1/2,L=0)$ in the experiment, especially in the heavy ion
collision.

\newpage

{\bf {Figure captions}}

\noindent
Fig.1. The total cross section of $\Lambda p$ elastic scattering.

\noindent
Fig.2. Energy level sketch for $(N\Omega)_{(2,1/2)}$ system.

\noindent
Fig.3. Energy level sketch for $(\Delta\Omega)_{(3,3/2)}$ system.

\noindent
Fig.4. S wave scattering phase shifts for $(N\Omega)_{(2,1/2)}$ system.

\noindent
Fig.5. S wave scattering phase shifts for $(\Delta\Omega)_{(3,3/2)}$ system.

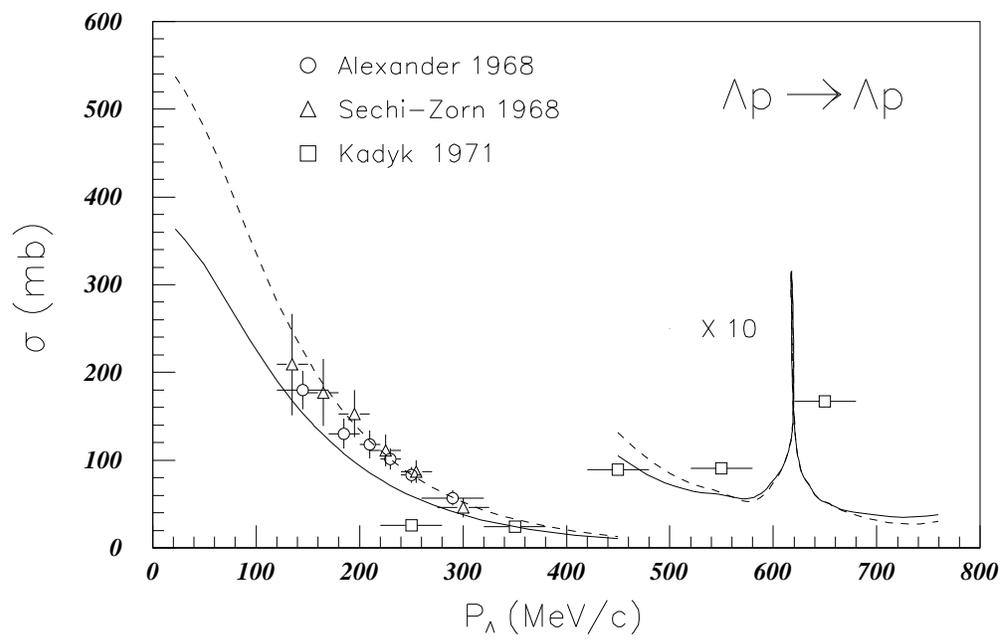
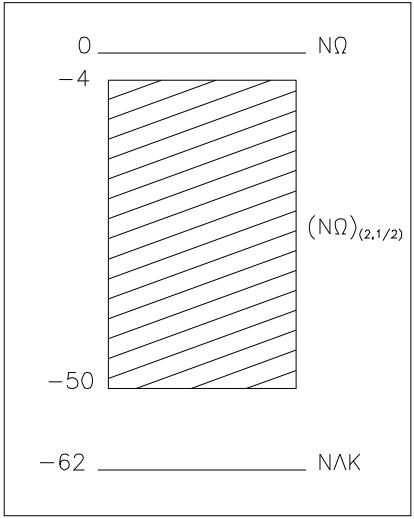
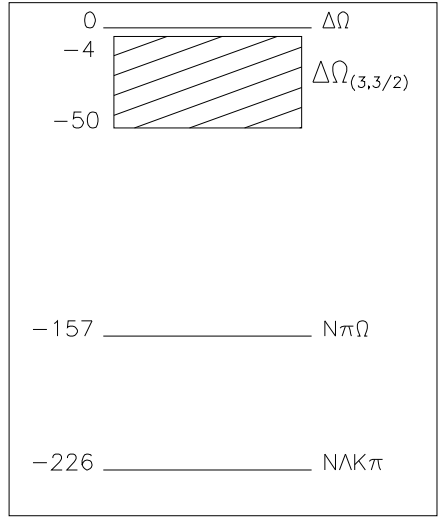
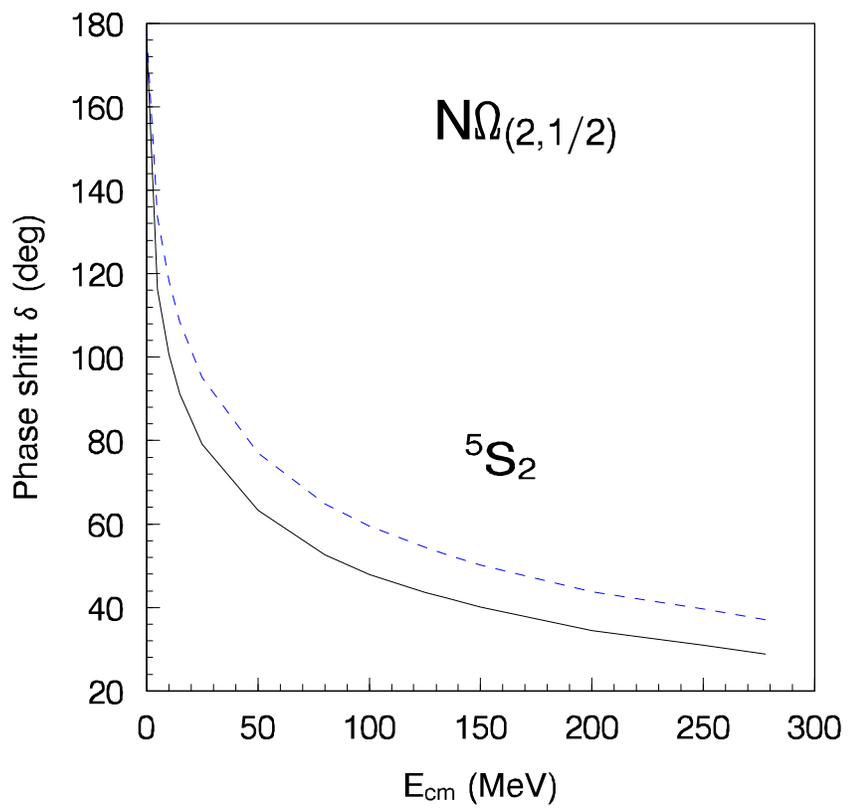
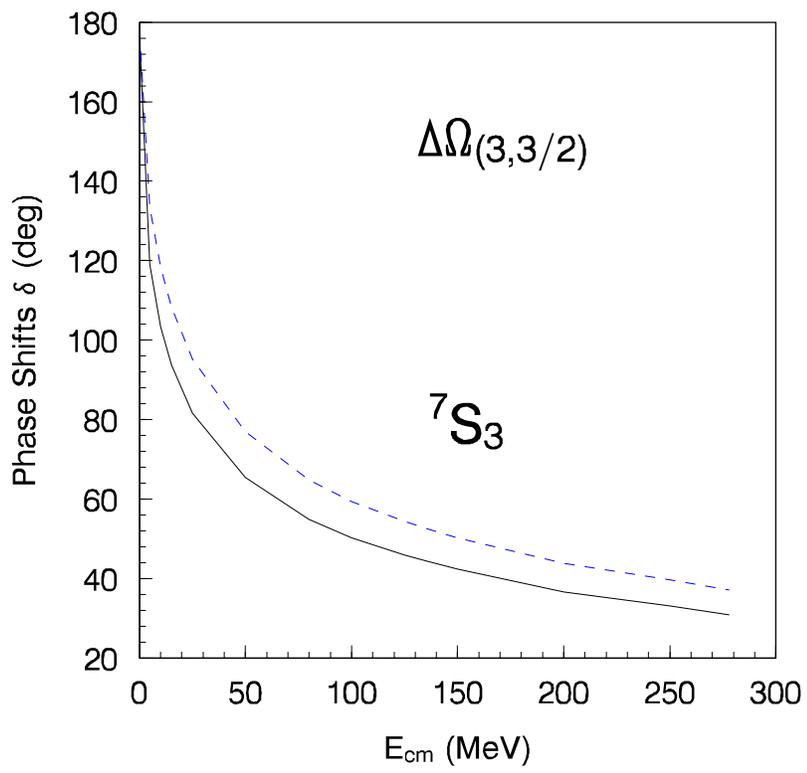
\end{document}